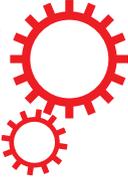

# SCIENTIFIC REPORTS

**OPEN** Overshoot during phenotypic switching of cancer cell populations

Alessandro L. Sellerio[1,2], Emilio Ciusani[3], Noa Bossel Ben-Moshe[4], Stefania Coco[5], Andrea Piccinini[5], Christopher R. Myers[6,7], James P. Sethna[6], Costanza Giampietro[8], Stefano Zapperi[1,2,9,10] & Caterina A. M. La Porta[8]

The dynamics of tumor cell populations is hotly debated: do populations derive hierarchically from a subpopulation of cancer stem cells (CSCs), or are stochastic transitions that mutate differentiated cancer cells to CSCs important? Here we argue that regulation must also be important. We sort human melanoma cells using three distinct cancer stem cell (CSC) markers — CXCR6, CD271 and ABCG2 — and observe that the fraction of non-CSC-marked cells first overshoots to a higher level and then returns to the level of unsorted cells. This clearly indicates that the CSC population is homeostatically regulated. Combining experimental measurements with theoretical modeling and numerical simulations, we show that the population dynamics of cancer cells is associated with a complex miRNA network regulating the Wnt and PI3K pathways. Hence phenotypic switching is not stochastic, but is tightly regulated by the balance between positive and negative cells in the population. Reducing the fraction of CSCs below a threshold triggers massive phenotypic switching, suggesting that a therapeutic strategy based on CSC eradication is unlikely to succeed.

The cancer stem cell (CSC) hypothesis suggests that tumors are organized in an aberrant cell hierarchy, in which differentiated cells have a limited capacity to proliferate and are produced by a subpopulation of parent CSCs that replicate indefinitely[1]. It is challenging to identify CSCs; human biopsies taken at specific clinical times cannot provide a complete tumor history, and animal tumor xenografts miss the physiological environment in which the tumor grows[2]. Despite these limitations, recent experiments *in vivo* have confirmed the presence of an aggressive CSC-like subpopulation in benign and malignant tumors[3–5].

The population dynamics of CSCs is, however, more complex than the strict hierarchy originally proposed. Non-CSCs breast cancer cells can revert to a stem-cell-like state even in the absence of mutations[6]. Similarly, in melanoma a small population of CSC-like JARID1B positive cells has been shown to be dynamically regulated in a way that differs from the standard hierarchical CSC model[7], reconciling earlier findings[8–10]. Microenvironmental factors, such as TGF$\beta$, are found to enhance the rate of switching from non-CSC cells to the CSC state[11]. This is in line with earlier results from our group showing that ABCG2 negative cells isolated from human melanoma biopsies reexpress this marker after a few generations *in vitro*[12]. The idea that the environment is able to induce cells to switch into a more aggressive phenotype was confirmed by Medema and Vermeulen, who explored limiting stemness by modifying microenvironmental factors known to support CSCs in tumors[13]. While mounting experimental evidence supports the switch from non-CSC cancer cells to the CSC state, the biological factors regulating this process are still unclear. Two possible scenarios have been invoked to solve this puzzle: the switch to the CSC state is either i) driven by genetic mutations or ii) regulated by epigenetic factors[14].

[1]Center for Complexity and Biosystems, Department of Physics, University of Milano, via Celoria 16, 20133 Milano, Italy. [2]CNR - Consiglio Nazionale delle Ricerche, Istituto per l'Energetica e le Interfasi, Via R. Cozzi 53, 20125 Milano, Italy. [3]Istituto Neurologico Carlo Besta, Milano, Italy. [4]Department of Physics of Complex Systems, Weizmann Institute of Science, Rehovot, Israel. [5]Dipartimento di Scienze Bomediche per la Salute, University of Milano, Milano, Italy. [6]Laboratory of Atomic and Solid State Physics, Physics Department, Cornell University, Ithaca, NY. [7]Institute of Biotechnology, Cornell University, Ithaca, NY. [8]Center for Complexity and Biosystems, Department of Bioscience, University of Milano, via Celoria 26, 20133 Milano, Italy. [9]ISI Foundation, Via Alassio 11C, Torino, Italy. [10]Department of Applied Physics, Aalto University, P.O. Box 14100, FIN-00076, Aalto, Finland. Correspondence and requests for materials should be addressed to C.A.M.L.P. (email: caterina.laporta@unimi.it)

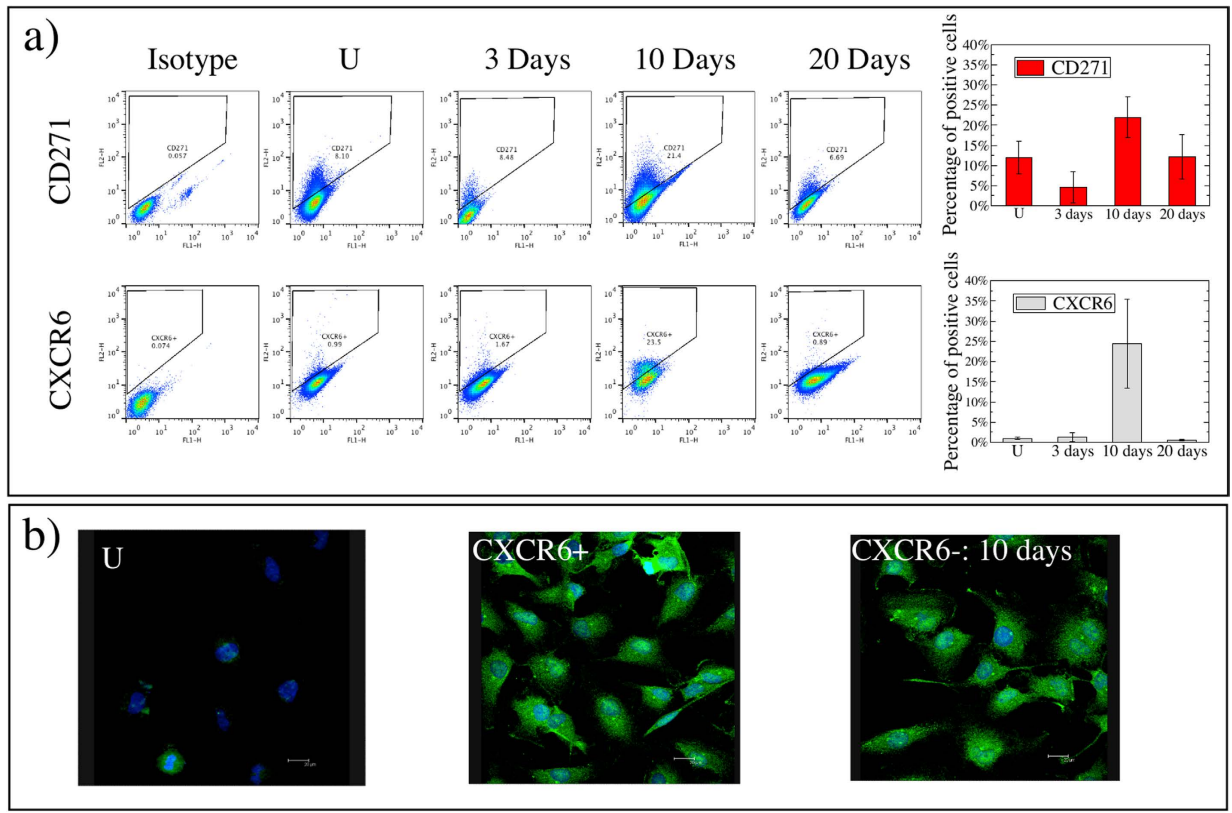

**Figure 1. Overshoot of CXCR6 and CD271 in human melanoma sorted cells.** CXCR6- or CD271-negative cells are sorted from IgR39 cells as described in Methods and plated under standard growth conditions. Cells are collected 3, 10 and 20 days after sorting, and analyzed by flow cytometry. Non-specific mouse IgG is used as isotype control (Isotype). The distribution of the markers in unsorted cells is also reported (U). Flow cytometry is performed using a FACSAria flow cytometer (Becton, Dickinson and Company, BD, Mountain View, CA). Data are analyzed using FlowJo software (Tree Star, Inc., San Carlos, CA). For each flow cytometry evaluation, a minimum of $5 \times 10^5$ cells are stained and at least 50000 events are collected and analyzed. (**a**) The level of expression of each markers in unsorted cells (U) and at different times after sorting is reported as flow cytometric analysis of a minimum of 5 independent experiments. (**b**) Unsorted cells (U), CXCR6-positive, and CXCR6-negative cells after 10 days sorting are fixed with 3.7% paraformaldeide and incubated with polyclonal anti-CXCR6 antibody (1:400, Abcam, Ab8023) overnight. Then the cells are incubated with the secondary antibody (anti rabbit Alexa488 1:250) for 1h and the nuclei counterstained with DAPI. The slides are mounted with Pro-long anti fade reagent (Invitrogen) and the images acquired with a Leika TCS NT confocal microscope.

In this paper, we report results indicating that the switching to the CSC state in human melanoma cells is regulated by an internal underdamped homeostatic mechanism controlling the fractional population, not dependent on external microenvironment. We measure the switching to the CSC state on the single-cell level with three biomarkers (CXCR6[15], ABCG2[12], and CD271[10]). Our group identified CXCR6 as biomarker of a CSC-like aggressive subpopulation in human melanoma cells[12,15]; it was also shown to play a critical role in stem cell biology, being linked to asymmetric cell division[15]. For all three markers, positive cells are known to give rise to a bigger tumor in immunodeficient mice than the negative ones[10,12,15]. The small tumor obtained with the negative population could be due to the use of an imperfect marker[16], to sorting errors or to phenotypic switching[6]. Here we follow the population dynamics of the negative subpopulation for longer times than earlier studies[6], recognizing that cells may require time to revert their phenotype (melanoma cells divide at a rate of ~0.60/day[17]). We observe that reducing the percentage of CSCs below a threshold induces cells to switch en masse, substantially overshooting the fraction seen in unsorted cells. This strongly suggests that a therapeutic strategy based on CSC eradication is unlikely to succeed.

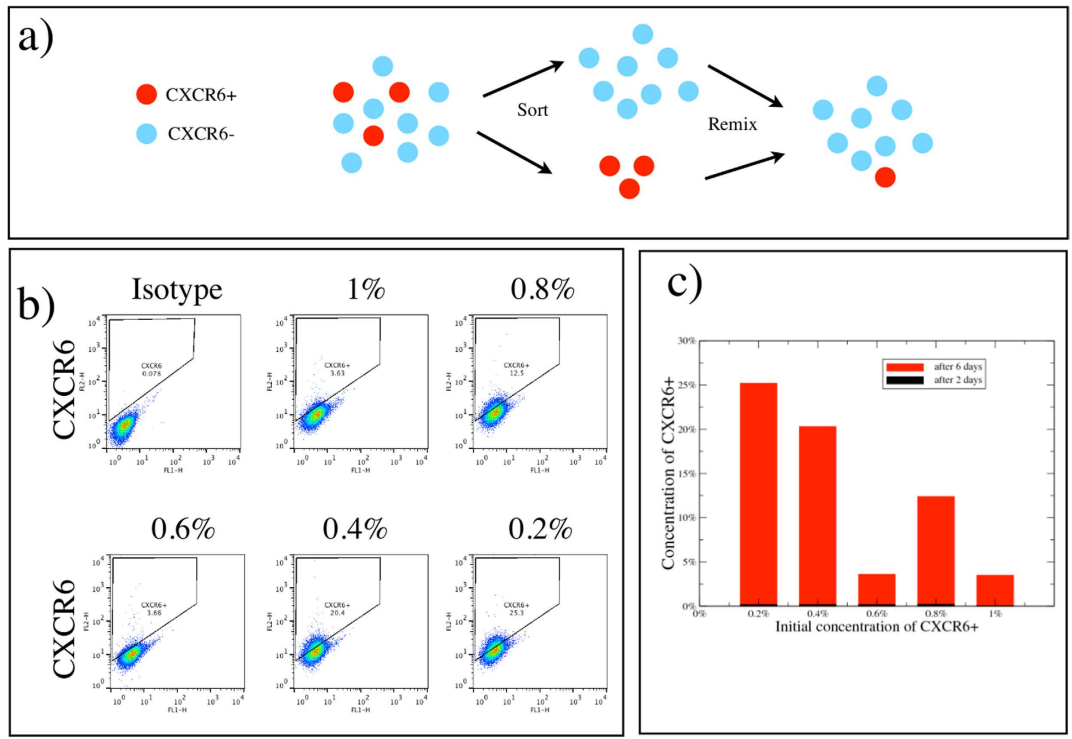

**Figure 2. The rate of phenotypic switching depends on the initial concentration of positive cells.** (**a**) CXCR6-positive and -negative IgR39 cells are sorted by flow cytometry and mixed with the fraction of positive cells varying from 0.2 to 1%. Cells are collected 2 and 6 days after sorting and the level of expression of CXCR6 is quantified by flow cytometry. (**b**) A representative example of flow cytometry for different initial fractions of CXCR6 positive cells. (**c**) The percentage of positive cells obtained in the different experimental conditions.

We then investigate the underlying regulatory mechanism. We provide evidence for the intriguing possibility that miRNAs play a key role in regulating the switch to the CSC state. miRNAs are small evolutionarily-conserved single-stranded RNAs which inhibit their target genes through binding to miRNA response elements (MREs)[18]. A single mRNA usually contains MREs for multiple miRNAs and, at the same time, an individual miRNA often targets up to 200 functionally diverse transcripts. miRNAs have been shown to suppress the expression of important cancer-related genes and have been proved useful in the diagnosis and treatment of cancer[19]. By studying the differential expression of miRNAs before and at the overshoot, we observe and characterize a complex miRNA regulatory network that plausibly governs the switching response.

## Results

### CXCR6, CD271 or ABCG2 negative human melanoma cells switch and overshoot after sorting.

IgR39 melanoma cells that are negative for CXCR6, ABCG2 or CD271 markers are sorted by flow cytometry, and then plated and grown under standard conditions. Considering that the positive cells represent only a very small fraction of the entire population, the possible error due to sorting is extremely low[16], as confirmed by the FACS analyses of the negative cells immediately after sorting (Fig. S1). Cells are collected 3, 10 and 20 days after sorting and analyzed for each marker by flow cytometry or immunofluorescence. Figure 1 shows the results for CD271 and CXCR6, while Fig. S2 reports the results for ABCG2. In all cases, the marker-negative cells re-express their marker in a time-dependent manner. In particular, the marker-positive population overshoots its initial level at 10 days post-sorting and then returns to the steady-state level at 20 days (Fig. 1 and Fig. S2). Immunofluorescence analysis confirms that CXCR6-negative cells after 10 days re-express such a factor at the same level as CXCR6-positive cells (Fig. 1). We next investigate whether the overshoot observed is dependent on the fraction of positive cells in the entire population. We thus sort by flow cytometry the CXCR6-positive cells and remix them with the negative ones in a fixed percentage ranging from 0.2% to 1% as shown in Fig. 2. A value around 1% represents the physiological condition of unsorted cells (Fig. 1). After 2 and 6 days, the remixed cells are analyzed by flow cytometry. Figure 2 demonstrates that mixed populations with a positive CSC fraction at or below 0.4% massively revert their phenotype after 6 days. We have previously shown that CXCR6-positive cells are also ABCG2-positive[15]. Herein we show that there is a strong overlap between the CXCR6-positive and CD271-positive populations (Fig. S3). This indicates that the observed effect is independent of the particular marker chosen to characterize CSCs, and in fact, the overshoot is observed for all markers.

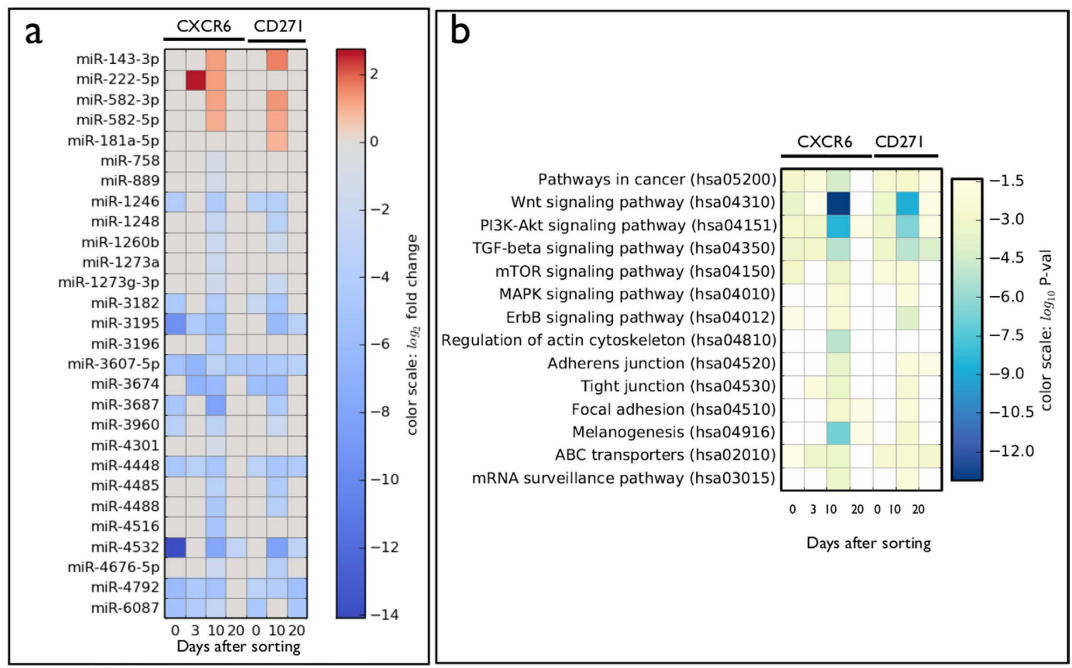

**Figure 3. Identification of differentially expressed miRNAs before and during the overshoot and the corresponding pathways involved.** (**a**) The color map shows miRNAs that are are differentially expressed with respect to unsorted cells for CXCR6- or CD271- negative cells at 3, 10 and 20 days after sorting. The cells are also analysed immediately after sorting (0). We filter by p-value ($<10^{-5}$), fold change($>2.0$), and absolute expression ($>0.01\%$). Increased levels are reported in red, decreased levels in blue. (**b**) For each condition in panel a, we identify the significantly affected pathways using the Diana-MirPath database as described in Methods. The colormap shows the estimated p-value for each pathway at different times. The p-value represents the probability that a given pathway is significantly enriched with the presence of mRNA targets of each miRNA[21].

**Short tandem repeat analysis of CXCR6 positive and negative human melanoma cells show no differences.** To check if CXCR6-positive and CXCR6-negative IgR39 cells display genetic differences, we analyze short tandem repeats (STR). These are microsatellites, consisting of a unit of two to thirteen nucleotides repeated hundreds of times in a row on the DNA strand, STR analysis measures the exact number of repeating units. A total of 23 STRs located on several chromosomes are analyzed as described in Methods. The results show no significant differences between sorted positive and negative cells in comparison to unsorted ones (Fig. S4 and S5).

**Differential expression of miRNAs before and during overshoot.** Human melanoma IgR39 cells express many miRNAs, as detailed in the supplementary data S1 (see Fig. S6). miRNAs influence most fundamental biological processes by ultimately altering the expression levels of proteins either through interference with mRNA translation or by reducing the stability of the mRNA in the cytoplasm[20]. Here, we consider how the miRNAs expression level evolves in time after sorting. As discussed in the Method section, we identify all the miRNAs that are significantly altered with respect to the unsorted condition for CXCR6- and CD271-marked cells after 3, 10 and 20 days. The results are reported in Fig. 3a in terms of their fold change with respect to the unsorted condition. First, note that the two subpopulations show the same trend, consistent with the fact that the CXCR6-positive subpopulation is also CD271-positive (Fig. S3). Second, note that more changes in miRNA expression appear at the overshoot at 10 days than at other times (Fig. 1). At 10 days, some miRNA levels increase in both marker populations (fold changes compared to the unsorted condition are reported in parentheses after miRNA names): miRNA 143—3p (5.6), miRNA 582—3p (4) and miRNA 582—5p (4). For the CXCR6-marked set only, we see an increase in miRNA 222—5p (2.4), while miRNA 181a-5p (4) increases only in the CD271-marked set. For both marker sets, most of the differential miRNA levels decrease (Fig. 3a). Before the overshoot, notable expression increases include miRNA 222—5p (6.5), with expression decreases for miRNAs 3607-5p, 3674 and 4448 decrease (1/90, 1/111 and 1/10, respectively).

Based on the assumption that miRNAs function by targeting complementary mRNAs, we hypothesize that differential changes in miRNA levels will produce corresponding changes in targeted mRNAs, and that mRNA targets can be predicted using bioinformatic tools that analyze miRNA-mRNA sequence complementarity. Using the Diana-MirPath database[21], we match differentially expressed miRNAs with pathways described in the KEGG database. We find that Wnt and PI3K-AkT signaling pathways are the most targeted by the miRNAs, as assessed by the p-value estimated by Diana-MirPath that accounts for both the number of targeted genes and the number of overall genes in each pathway. Figure 2b displays the targeted pathways in terms of their statistical significance (p-value), while Fig. S7 reports the number of miRNAs targeting each pathway, and the number of targeted genes in each pathway. In Fig. 3a,c, we illustrate the network of interactions between the differentially expressed miRNAs before and at the overshoot in CXCR6-negative cells, respectively; Fig. 3b,d restrict these networks to highlight the Wnt and PI3K-Akt pathways identified in Fig. 3 as being especially significant. The interaction network is quite complex, displaying a remarkable degree of redundancy, with many miRNAs targeting the same pathway, and pleiotropy, with each miRNA targeting multiple genes and pathways. Interestingly, the level of miRNA 222—5p, which is identified as targeting the Wnt gene, increases significantly before the overshoot and then decreases at the overshoot.

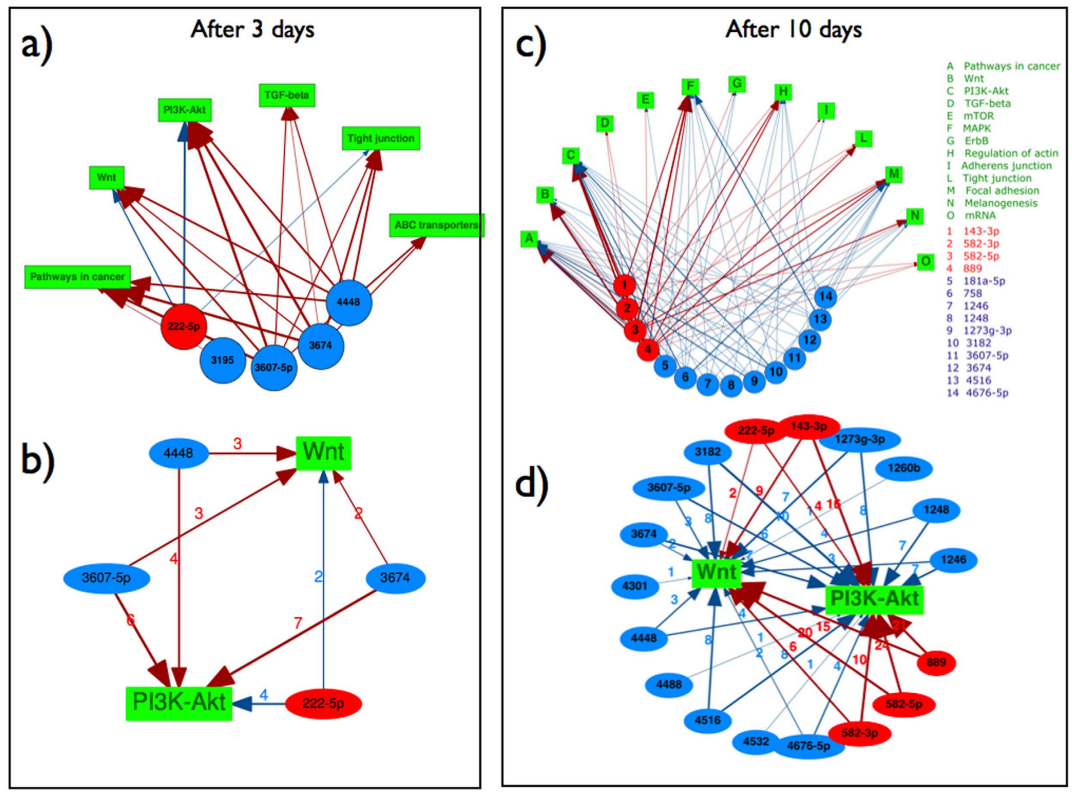

**Figure 4. Interaction network between miRNAs and pathways.** (**a**) The network formed by differentially expressed miRNAs in CXCR6-negative cells before the overshoot (3 days after sorting) and pathways identified to be significantly targeted by those miRNAs. (**b**) The same network as in a) restricted to Wnt and PI3K signalling pathways. (**c**) The network formed by differentially expressed miRNAs in CXCR6-negative cells at the overshoot (10 days after sorting) and pathways identified to be significantly targeted by those miRNAs. (**d**) The same network as in (**c**) restricted to Wnt and PI3K signalling pathways. For all panels the miRNAs are shown in red when they increase and in blue when they decrease; the arrows are red when they induce an increase of the targeted pathway and in blue when they induce a decrease.

**Dynamic regulation of $\beta$catenin before and during overshoot.** In order to experimentally validate the bioinformatic analysis suggesting that the Wnt signalling pathway is one of most significantly targeted by miRNAs before and at the overshoot, we investigate the dynamic regulation of $\beta$catenin and other related factors such as Adenomatous Polyposis Coli (APC), Axin, and Glycogen Synthase Kinase 3$\beta$ (GSK3$\beta$). We restrict our analysis to the CXCR6-sorted population, because CXCR6-negative cells express the same miRNAs as CD271-negative ones (Fig. 3). In the absence of Wnt signals, $\beta$catenin is found in two distinct multiprotein complexes: one complex, located at the plasma membrane, couples cadherins with the actin cytoskeleton, whereas the other complex containing the proteins APC, Axin, and GSK3$\beta$ targets $\beta$catenin for degradation. Wnt antagonizes the APC—-Axin—-GSK3$\beta$ complex, resulting in an increase in the pool of free cytoplasmic $\beta$catenin. Stabilized $\beta$catenin translocates to the nucleus, binds LEF1/TCF4 factors and activates Wnt target genes, a process that may involve the disruption of interactions between LEF1/TCF4 and corepressor proteins. However, The LEF1/TCF4 transcription factors can also interact with other cofactors and play an architectural role in the assembly of multiprotein enhancer complexes, which may allow for the integration of multiple signalling pathways[22]. Using real time PCR, we quantify the level of expression of mRNA of $\beta$catenin, Axin, APC, GSK3$\beta$, LEF1, TCF4 as shown in Fig. 5a. After a significant decrease of all these factors 3 days after sorting, they all increase at the overshoot. Nevertheless, western blots show that $\beta$catenin is not more activated at the overshoot (Fig. 5b). These results are in agreement with the intracellular localization of $\beta$catenin detected by immunofluorescence. In particular, $\beta$catenin is localized at the cell membrane, 10 days after sorting, while it is more homogeneously distributed in unsorted cells (Fig. 5c). In addition, these cells express N-cadherin[12] whose expression also changes during switching (Fig. 5a). Finally, downstream factors also regulated by $\beta$catenin, such as c—myc and cyclin D1 are also affected: before the overshoot cyclin D1 increases and N-cadherin decreases. At the overshoot, cyclin D1 decreases and N-cadherin increases.

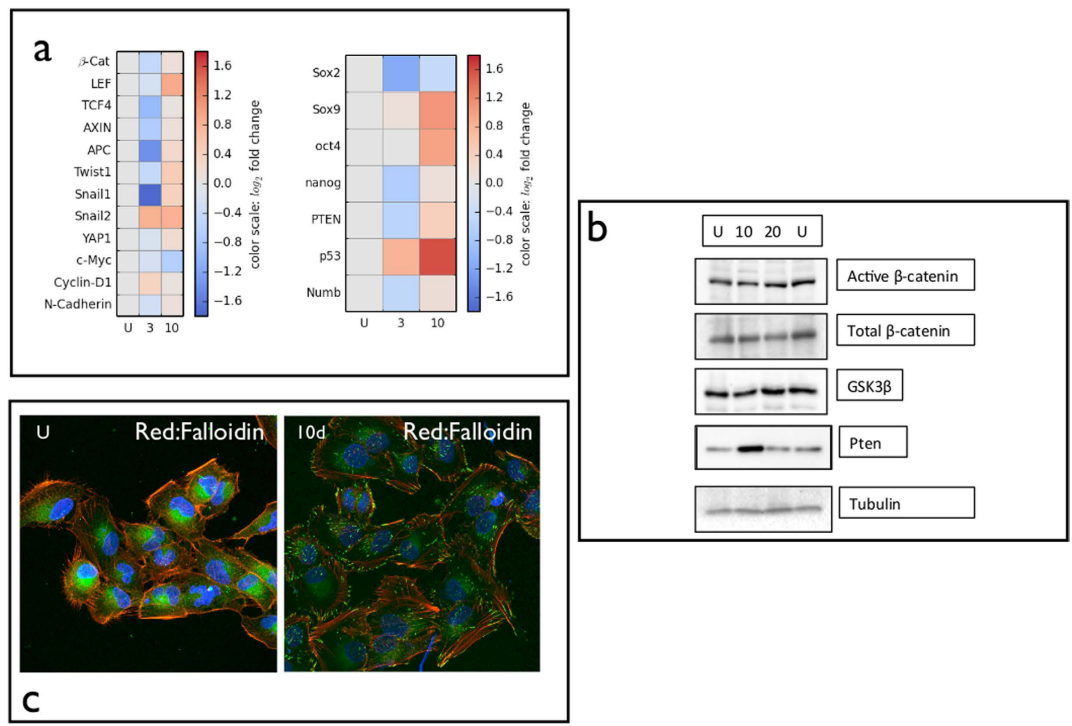

**Figure 5. Expression of regulatory factors during phenotyping switching.** (**a**) Colormap of real time PCR has been carried out on unsorted cells and CXCR6-negative cells 3 and 10 days after sorting as described in Methods. The colormap shows the log of fold change of each factor at different times after sorting. (**b**) Western blot of unsorted cells and CXCR6-negative cells 10 and 20 days after sorting as described in Methods. Tubulin has been used as housekeeping gene. (**c**) Immunofluorescence of $\beta$catenin (green) and Falloidin (red) of unsorted cells (US) and CXCR6-negative cells 10 days after sorting (10 days).

**Dynamic regulation of endothelial-mesenchimal transition factors before and during overshoot.**
We characterize CXCR6 phenotypic switching cells for the most important transcription factors playing a critical role in endothelial mesenchimal transition (EMT). This process promotes self-renewal capabilities and the stem-like phenotype and is correlated to a poor prognosis of solid tumors[23,24]. Twist and Snail as well as Yap, which regulate the Hippo tumor suppressor pathway[25], and Numb, a CSC marker for breast cancer[26,27] all increase at the overshoot (Fig. 5a). Numb has been shown recently to play as a regulator of p53 preventing its ubiquitination and degradation[26]. Accordingly, during switching we observe an increase in the mRNA level of p53 (Fig. 5a). Under the same conditions, Sox9 and Oct4 levels also increase (Fig. 5a). This appears to be in agreement with the role of Sox9 in adult stem cell maintenance[28] and in melanoma[29]. In contrast, Nanog does not change and Sox2 decreases (4a). PTEN (phosphatase and tensin homologue) is another factor that plays a critical role in stem cell maintenance and its loss is involved in the development of CSCs[30]. As shown in Fig. 5a,d, PTEN first decreases when the level of the CSC markers is still low, and then increases at the overshoot. This is consistent with the role of PTEN for CSCs[30]: 3 days after sorting, when CSC are starting to switch, the level of PTEN is low, thus promoting CSC development. At the overshoot, when the population CSCs is large, PTEN is high leading to a subsequent reduction of the CSC population. Before the overshoot, the cells show a decrease of all EMT-related factors with the exception of Snail2 and p53 (Fig. 5a).

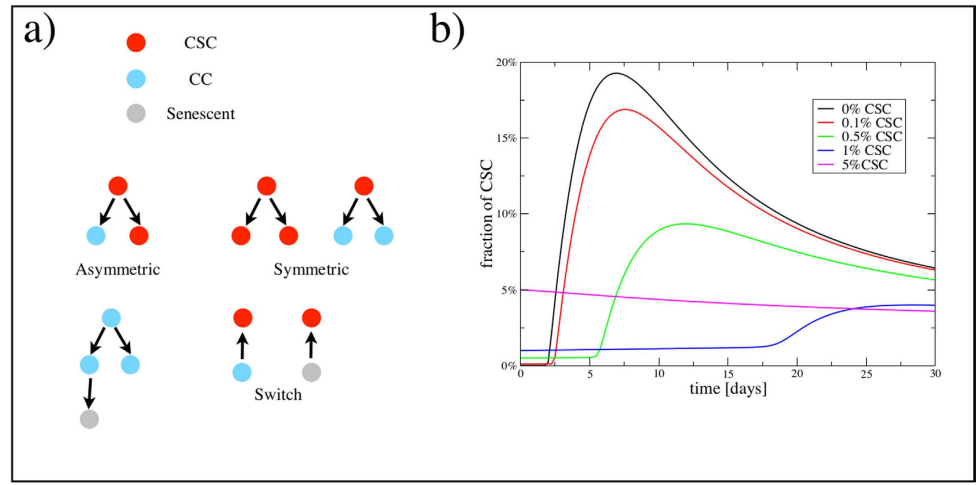

**Figure 6. Mathematical model explains CSC overshoot in phenotypic switching.** (**a**) A schematic representation of the model considering CSC, cancer cells (CC) and senescent cells, with a possibility of phenotypic switching depending on the level of miRNA. (**b**) Simulations of the model for cell populations with different initial fraction of CSC display an overshoot in the fraction CSC for sufficiently low initial concentrations, in agreement with experiments.

**miRNA—222 silencing in CXCR6 negative cells during switching.** To confirm the effect of miRNA—222 on the Wnt and PI3K-pathways, we silenced the unsorted and the marker-negative cells with Validated AMBION SILENCER Select siRNA 222. As control, siRNA targeted GAPDH has been used. As shown in Fig. S9, the silencing of miRNA222 in unsorted cells, leads to a specific change in a set of genes linked to Wnt and PI3K-pathways. To confirm the specificity of silencing, the silencing of the housekeeping gene GAPDH does not give the same pattern of siRNA222 but silences GAPDH only (Fig. S8). In sorted CXCR6 cells the silencing of miRNA—222 for 3 days affects the Wnt and PI3K pathways. In particular $\beta$-catenin is up-regulated as expected.

**Population dynamics model miRNA-dependent is consistent with overshoot in phenotypic switching.** To understand the observed CSC population dynamics in which the CSC-like cell population overshoots after sorting, we consider a simple mathematical model for the dynamics of hierarchical cell populations with CSCs, cancer cells and senescent cells (Fig. 6a)[16,31]. In the absence of phenotypic switching, CSCs give rise to cancer cells and the reverse is not true[31]. Hence a perfectly sorted CSC-negative population does not express CSC markers again. If we introduce a small probability for cancer cells to switch back to the CSC state, CSCs will eventually reappear in the negative population but no overshoot would be observed[6]. We explain the presence of an overshoot assuming that the switching probability is usually zero but increases drastically after the activation of a miRNA network which responds to the lack of CSCs in the population. This mechanism is formalized in a mathematical model described in the supplement. We perform numerical simulations of the model, sorting and remixing CSCs and non-CSC cells as in Fig. 2. We then follow the dynamics of the CSC population, which displays an overshoot whose amplitude decreases as the initial fraction of CSCs increases (Fig. 6b).

**CSCs in human breast cancer display related miRNA and transcriptome signatures.** The identification of specific signatures for each tumor, and for tumor development to more aggressive conditions, is a mainstream goal. Our data highlight the role of a set of genes regulating the transition between cancer cells to CSC in human melanoma. To assess whether these results might have implications for other cancers, we analyze the level of expression of the implicated set of genes in other kinds of human tumors such as breast cancer (METABRIC dataset; see Methods). First, we find that miRNA-222 is more expressed in tumors that display the Embryonic Stem Cell (ESC) signature (fold change = 1.12, p-value = 0.002). Second, several important mRNAs that are modified in melanoma at the overshoot are seen to vary in ESC-like breast tumors (see Fig. 7a). Finally, we examined miRNAs that are significantly differentially expressed between ESC and non-ESC tumors, and we find that predicted mRNA targets of these miRNAs are involved in a set of relevant pathways, including Wnt and PI3K-AkT (Fig. 7b).

### Discussion
The CSC hypothesis has been partially confirmed, despite the complex challenges of biomarker identification and the artificial environments presented by xenografts[16]. Three recent and independent papers identified and validated the existence of a hierarchically organized aggressive subpopulation that sustains tumor growth in mouse models of brain[5], skin[4] and intestinal[3] tumors: CSCs indeed exist and arise *de novo* during tumor formation in intact organs. However, the boundary between CSC cells and ordinary cancer cells appears more porous than originally conceived: several groups have shown that cancer cells can revert to the CSC state, although the biological mechanism leading to this phenotypic switching remained unclear[6,11,32].

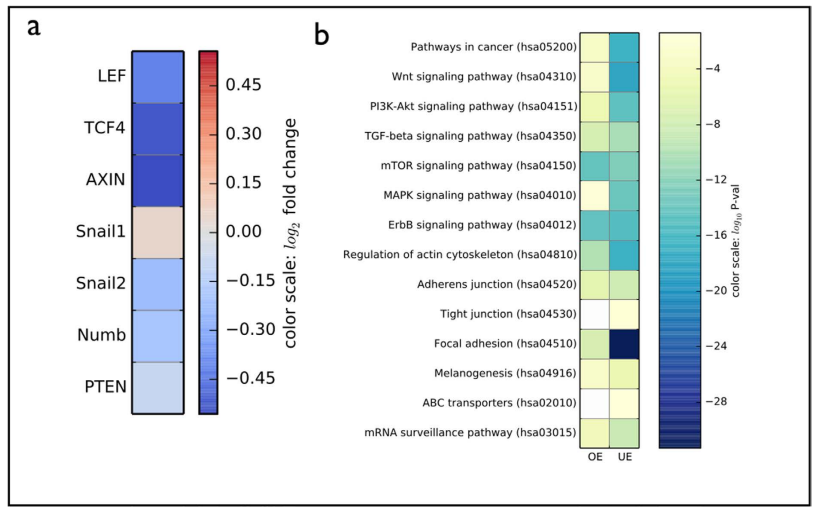

**Figure 7. mRNA and miRNA expression in breast tumors with ESC signature.** mRNA expression data from 1980 breast tumors and miRNA expression data from 1283 breast tumors[47,55] are separated in two classes depending on the presence of an ESC signature[56]. (**a**) We report a set, corresponding to the factors in Fig. 5a, of significantly changed mRNA between ESC and non-ESC tumors. The color represents the relative fold-change of each mRNA for the two sets of tumors. (**b**) A large set of miRNA are significantly over-expressed (OE) or under-expressed (UE) between ESC and non-ESC tumors. We report the pathways putatively targeted by these two sets of miRNAs. The color represent the p-value. A low p-value suggests that the pathway is likely to be targeted by the set of OE or UE miRNAs.

Here we provide clear evidence for an environmentally triggered, homeostatic population regulatory mechanism, controlled by epigenetic changes mediated by miRNA expression, governing the phenotypic switching in melanoma cell lines back to the CSC state. We do so using a multifaceted approach, combining the use of CSC-specific markers and single-cell sorting to trigger switching, miRNA analysis of the concurrent inter-cell regulatory activity, and mathematical analysis of the resulting population shifts.

To segregate CSCs from the non-CSC population, we use three markers validated for human melanoma cells: CXCR6 (which also regulates the switch between asymmetric to symmetric cell division[15]), CD271[10] and ABCG2[12]. For all three markers, ten days after sorted negative cells are re-plated they display a significant overshoot in the re-expression of the marker. Furthermore, the amplitude of the overshoot is dependent on the percentage of positive cells present in the population, suggesting that the balance between positive and negative subpopulation is a factor in triggering phenotypic switching. When the number of positive cells in the population falls below a threshold, negative cells switch and become positive (CSC). This mechanism acts as an underdamped homeostatic mechanism, compensating for the depletion of CSCs and returning the population to the initial steady state. The fact that we find similar overshoots for the three markers is evidence both that the initial sorting is valid (keeping the same non-CSC subpopulation) and that the overshoot is multifaceted (reverting broadly to the original CSC phenotype). In addition, we show that CXCR6 positive and negative cells share the same STR profiles, suggesting that they do not differ genetically. The overshoot and the lack of significant mutations indicates that the switching is not a statistically random event, but is regulated by the CSC population.

To explore the regulatory mechanism, to show that it involves the tumor population as a whole (and not a hyperstimulated mis-sorted CSC population), and to definitively show that the CSC switching is due to intrinsic factors released by the cells rather than extrinsic factors, we study the miRNA expression. We focus on miRNAs as a possible way used by cells to communicate between each other and to produce the factors needed to switch. miRNAs are epigenetic factors that control crucial aspects of cellular life such as duplication, differentiation and stem-cell-like phenotypes. It is known that even small differences in miRNA expression levels can have profound functional consequences for fundamental cellular programs[33]. Since we are not aiming to validate a specific miRNA for melanoma, we use a well-characterized typical human melanoma cell line, which importantly expresses high levels of CSC markers and originates from a primary tumor[15,17,31]. Evidence of phenotypic switching in different melanoma cell lines has been reported by our group and others (WM115[12] and WM3734[7]). The basal miRNA expression in cell line used here is high, suggesting an intense regulatory activity and strong interactions between cells.

Before and after the overshoot, we identify all the miRNAs that are differentially expressed compared to the unsorted condition. The results are largely independent of the marker used for sorting, consistent with the co-expression of CXCR6 with the other markers. Based on a bioinformatic analysis, we observe that miRNAs act in a pleiotropic and redundant manner affecting many intracellular pathways and therefore creating an intricate network of connections. In spite of the complexity of the network, two main pathways appear to be primarily targeted by the differentially expressed miRNAs: the Wnt and PI3K signalling pathways. Before the overshoot, miRNA 222—5p (targeting Wnt), increases, contributing to a decrease of the activity in the Wnt-pathway and in the level of $\beta$catenin. The $\beta$catenin level in the plasma membrane is indeed clearly shifted at the overshoot. This is consistent with the observation of an interesting recent paper on embryonic stem cells, where the transcriptional activity of $\beta$catenin under self-renewal conditions was shown to be negligible due to its membrane localization and the subsequent complex formation with Oct4[34]. We also compared the behavior of CXCR6 negative cells with that of unsorted cells to silencing of miRNA—222. In both cases silencing led to the expected increase in $\beta$catenin. But the unsorted cells also showed an increase of LEF1, TCF4, Axin, Twist, Snails and p53, an increase not observed in CXCR6-negative cells. Thus the response of negative and unsorted cells to silencing this stimulated pathway also differs.

The main function of PTEN is to negatively regulate the PI3K-Akt pathway. The loss of PTEN results in an overactive Akt, which induces proliferation and promotes survival by inhibiting apoptosis[35]. Accordingly, in hematopoietic stem cells cooperation between PTEN and Wnt pathways occurs during phenotypic switching[36]. Recently, PTEN-null melanocyte stem cells have been shown to be resistant to exhaustion after repeated depletions, a phenotype that is similar to the one observed in PTEN-null neural stem cells[37]. Interestingly, PTEN is regulated during phenotypic switching and increases substantially, together with p53, when the cells are re-expressing the CSC phenotype. It is well known that PTEN can regulate p53 protein levels and its transcriptional activity[38]. A dynamic regulation of Wnt and PTEN-p53 pathways occurs and regulates the phenotype of negative cells due to the differentially expressed miRNAs: before the overshoot, when negative cells are about to switch, PTEN and $\beta$catenin are low, while at the overshoot PTEN and p53 increase, contributing to differentiation and thus decreasing the number of CSCs to their steady state level. This is in line with a recent review discussing the role of p53 as the barrier to CSC formation: p53 is necessary to maintain a pool of normal stem cells by controlling their quantity and quality[39]. These properties are still present in CSCs.

Another intriguing aspect of our experimental results is the relation between CSC switching and the epithelial—mesenchymal transition (EMT), a profound event for large-scale cell movement during morphogenesis at the time of embryonic development. During this process, epithelial cells lose contact with their neighbors and gain mesenchymal properties, enabling them to break through the basement membrane that separates different tissues within the embryo[40]. Because a similar process has been observed at the invasive front of metastatic cancer, the EMT is considered to be a hallmark of neoplastic transformation. The EMT results from the induction of transcription factors that alter gene expression to promote loss of cell—cell adhesion, leading to a shift in cytoskeletal dynamics and a change from epithelial morphology and physiology to the mesenchymal phenotype. Before the overshoot, all the EMT factors decrease with the exception of Snail2. Emerging evidence indicates that Snail confers to CSC-like traits to tumor cells, promoting drug resistance, tumor recurrence and metastasis[41]. Moreover, direct regulation of the Snail promoter is due to different growth factors, such as TGF—$\beta$[42]. The latter is one of the pathways targeted by miRNAs before the overshoot. On the other hand, the expression of Snail appears to be closely related to metastasis[41], in accordance with the increased level of CSCs at the overshoot. Twist, also involved in the EMT, is also increased at the overshoot[43]. However, at the overshoot (when the cells are expressing PTEN and p53) the levels of cyclin D1 and c—myc decrease whereas SOX9 (shown to be expressed in more differentiated melanoma phenotypes) increases[29].

In order to understand how an overshoot in the expression of CSC markers can possibly arise after sorting, we turn to a mathematical model for the population dynamics of cancer cells. Previous models for phenotypic switching introduce a small probability for cancer cells to switch back to the CSC state[6,16]. Such models predict that the CSC population will re-approach its steady-state concentration after sorting, but without an overshoot. In our model, the switch probability is dependent on the expression level of a set of relevant miRNAs which is in turn controlled by the population of CSCs. When the CSC population is depleted, the miRNA expression level shoots up, triggering a large switching probability. This induces an overshoot in the fraction of CSCs in the population that then falls back to the steady state level as the level of miRNAs decreases. The model illustrates in simple terms how cancer cells react to the depletion of CSCs, showing that phenotypic switching is not just a random event, as assumed in previous works[6], but an active response to a depletion of the CSC population.

A similar protective mechanism might be at play in stem cells niches: without switching, our model is very similar to a stochastic model for stem cell evolution that has been found to describe accurately the distribution of clone size *in vivo*[44]. In the case of stem cells, however, tissue homeostasis requires that, on average, each stem cell gives rise to one stem cell (i.e., the parameter $\varepsilon$ is equal to zero[31]). In these conditions, the model predicts that the probability of extinction for a clone originating from a single stem-cell tends to one in the long-time limit[45] but when the stem cell pool is large enough, fluctuation-induced extinction is very unlikely[44]. If an external perturbation suddenly reduces the stem cell pool, however, the model predicts that extinction becomes likely. Phenotypic switching represents an effective response to this threat.

More broadly, (non-cancerous) stem cells are thought to be subject to two homeostatic regulatory mechanisms[46]. As above, the first keeps the stem cell population near a fixed number. We model this first homeostasis mechanism in CSCs not by regulating the probability of symmetric divisions into CSCs, but by introducing a regulated switch into the CSC state. This choice is suggested by the concentration dependence of the overshoot; a more complete depletion of CSCs would take longer to recover in a model of regulated CSC symmetric divisions, but would respond more quickly in a switching model — as observed in our experiments (Fig. 2c). The second homeostatic mechanism in stem cells is thought to regulate the generation of new differentiated cells to control their number (e.g., organ size). Perhaps the cancer cell lines we study have lost the second control mechanism, but retain the first? All together, our data show that the number of CSCs is controlled by a complex network of miRNAs that regulate in particular the Wnt and PTEN pathways and EMT-related factors. We show that a similar set of genes is expressed in METABRIC ESC-like human breast tumors[47] with respect to non ESC, demonstrating that this signature is not specific to melanoma. Interestingly, in the same dataset miRNA—222 is significantly increased.

A direct consequence of our findings is that reducing the number of CSCs below a threshold induces the cells to switch. Rather than targeting and decreasing the CSCs as the 'root' of the cancer, our findings suggest that a balanced reduction in both populations is needed to keep the tumor in a state of asymmetric cell division as it is reduced. In stem cells the locality of Wnt signalling dictates differentiation and spatial confinement in a niche[48]. Our results suggest that CSCs and their offspring exhibit the presumed stem cell capability of regulating the number of stem cells in the niche and thus avoid extinction. We are not, however, aware of a switching mechanism from differentiated cells to stem cells being used to achieve this end in stem cell niches. A practical consequence of these findings is that since IgR39 cells are heterozygous for the BRAF V600 mutation (Fig. S9), a drug acting on such a factor should affect both CSCs and cancer cells. It could thus shrink the tumor at short times but lead to a relapse of the tumor at longer times due to the depletion of CSCs, inducing a switch in the population. This is in agreement with a recent paper where a minor subset of melanoma cells, required for tumor maintenance and expressing high levels of Jarid1B, are shown to undergo a marked expansion when treated with either BRAF inhibitors or cytotoxic chemotherapeutic drugs[7]. At least for melanoma, the best strategy seems to be to find a balance for the percentage of CSCs so that it does not fall below threshold, or a complete resection of the tumor. In addition, miRNA-222 is a promising target for suppressing phenotypic-switching replenishment in CSC-depleted tumors, not only in melanoma but also in breast cancer.

## Methods

**Cell lines.** Human IGR39 cells are obtained from Deutsche Sammlung von Mikroorganismen und Zellkulturen GmbH and cultured as previously described (standard conditions)[15].

**Flow Cytometry.** Cells are analyzed or sorted for phycoerythrin (PE) anti-human CXCR6 (R& D System, Minneapolis, MN), PE anti human CD271 (BD Pharmingen), fluoroscein isothiocyanate (FITC) anti-human ABCG2 (R& D Systems, Minneapolis, MN), FITC anti human fascin (Abcam, ab126772). Indirect flow cytometry is carried out with anti human p53 (Abcam Ab156030), anti human MKAC (Abcam, Ab187652) or anti human PTEN (Abcam, Ab32199). Secondary FITC antibody is used for indirect florescent (1:250, Alexa Fluo 488). All samples are analyzed using one flow cytometry with non-specific mouse IgG used as isotype controls. For each flow cytometry evaluation, a minimum of $5 \times 10^5$ cells were stained and at least 50000 events were collected and analyzed ($10^6$ cells were stained for sorting). Flow cytometry sorting and analysis was performed using a FACSAria flow cytometer (Becton, Dickinson and Company, BD, Mountain View, CA). Data were analyzed using FlowJo software (Tree Star, Inc., San Carlos, CA). For double staining with CXCR6 and CD271 markers, the cells are incubated with FITC anti-human CXCR6 (R& D System, Minneapolis, MN) and PE anti human CD271 (BD Pharmingen).

**Short tandem repeat (STR) analysis.** DNA is extracted by DNA MiniPrep kit (GN170, Sigma) and amplified using the commercial kits NGMSelect (Applied Biosystems)[49] and Powerplex 16[50] (Promega) following the manufacturers' instructions. The amplified loci are: D10S1248, VWA, D16S539, D2S1338, D8S1179, D21S11, D18S51, D22S1045, D19S433, TH01, FGA, D2S441, D3S1358, D1S1656, D12S391, SE33+ Amelogenin (NGMSelect) and D3S1358, TH01, D21S11, D18S51, PENTA E, D5S818, D13S317, D7S820, D16S539, CSF1PO, PENTA D, VWA, D8S1179, TPOX, FGA+ Amelogenin (Powerplex 16). The chromosomal location of the analysed STRs is displayed in Tables S1 and S2 and Figs S4 and S5. The resolution of the amplified fragments was carried out on ABI 310 Genetic Analyzer, followed by Genescan 3.1.2 and Genotyper 2.1 or GeneMapper 3.2 analysis. All the procedure of STR typing was repeated at least twice for each sample.

**BRAF V6oo mutation.** Using standard PCR conditions, BRAF exon 15 was amplified using the protocol of Davies *et al.*[51].

**Immunofluorescence.** Subconfluent cells grown on glass coverslips are fixed with 3.7% parafolmaldeide in PBS for 10 min, permealized with 0.5% Triton X-100 in PBS for 5 min at room temperature and incubated with 10% goat serum in PBS for 1 hr. The cells are stained with polyclonal anti-CXCR6 antibody (1:400, Abcam, Ab8023) or anti-$\beta$-catenin (MIllipore, MAB2081, 1:200 nostro; anti-Active $\beta$Catenin (Anti-ABC) Antibody, Millipore, 05-665, 1:50) or Anti-Fascin antibody (DyLight 488, ab156578 Abcam) overnight at 4 °C. Thus, after a brief washing with PBS, CXCR6 staining cells are incubated with the secondary antibody (anti rabbit Alexa488 1:250) for 1 h. The cytoskeleton is stained with Falloidin (Acti-stain 555 Fluorescent Phalloidin, Cytoskeleton Inc, 100 nM) at room temperature for 45 min. The nuclei are counterstained with DAPI and the slides mounted with Pro-long anti fade reagent (Life technologies). The images are acquired with a Leika TCS NT confocal microscope.

**Isolation and detection of miRNAs.** miRNeasy Micro kit (Qiagen, 217084) has been used for the purification of total RNA including miRNA according to the manufacturer's instructions. The integrity of total RNA samples is checked by Agilent Technologies Bioanalyzer 2100 through calculation of the RNA Integrity Number (RIN). Samples with RIN greater or equal to 8 are used for library preparation of total RNAs. The Illumina TruSeq Small RNA Sample Preparation kit is used to prepare libraries of miRNA suitable for sequencing. The sample preparation is performed using 1 $\mu$g of input total RNA. The first step of the sample preparation consisted in ligating sequences ("adapters") to the miRNAs. The protocol takes advantage of the natural structure common to most known miRNA molecules. Most mature miRNAs have a 5$'$-phosphate and a 3$'$-hydroxyl group as a result of the cellular pathway used to create them. The Illumina adapters in this kit are directly, and specifically, ligated to miRNAs. After adapter ligation, reverse transcription followed by PCR was used to create cDNA constructs based on the small RNA ligated with 3$'$ and 5$'$ adapters. This process selectively enriches those fragments that have adapter molecules on both ends. PCR is performed with two primers that anneal to the ends of the adapters. In one of two primers a sequence "tag" is present (index sequence). Each sample is added with a unique index sequence. The amplified libraries are checked on Bioanalyzer 2100 to verify the presence of the peaks corresponding to the miRNAs (peaks range from 145 bp to 158 bp). In Fig. S6 we show an example of the electropherogram of one of the amplified libraries. The checked amplified libraries are run on 6% Novex TBE Gel and the 147 nt band containing mature miRNA and the 157 nt band containing piwi-interacting RNAs, as well as some miRNAs and other regulatory small RNA molecules, are selected from the gel, purified and checked on Bioanalyzer chip. In Supplementary Fig. 2 a final library ready to be sequenced is reported. The 12 libraries are pooled together and the pool is sequenced on Illumina platform. Sequencing of the pool is performed on Illumina HiSeq1500 platform in single read protocol with the production of sequences of 50 bp in length. The number of reads produced per sample ranges from 6.8 million and 12.3 million. The percentage of $\geq$Q30 Bases (PF) and Mean Quality Score (PF) are two parameters used to estimate the quality of the sequencing run and of the produced sequences. Both the percentage of Q30 Bases (PF) and Mean Quality Score (PF) are higher ($>$98.16 and $>$38, respectively), suggesting a run of good quality.

**miRNAs analysis.** The FASTQ sequences were trimmed to remove the adapter sequences using the Cutadapt software toolbox[52]. The correctness of trimming and the initial sequence quality assessment was made using the combined software toolboxes TrimGalore! (http://www.bioinformatics.babraham.ac.uk/projects/trim_galore/) and FastQC (http://www.bioinformatics.babraham.ac.uk/projects/fastqc/). Data indicate high signal to noise ratio; no spurious sequences are detected in any of the samples. We counted the unique trimmed sequences and, for each sample, generated files listing all the unique sequences with the corresponding number of occurrences. The sequence occurrence files were uploaded to the Miranalyzer online tool[53] as input data; Miranalizer was run, using the default parameters, singularly on all 12 data sets to identify predicted miRNA sequences. We compared the unsorted conditions (US) with all the other conditions selecting candidate miRNAs by filtering according to p-value ($<10^{-5}$), fold change($>$2.0), and absolute expression($>$0.01%). This allows us to concentrate on good candidates (low p-value) that have undergone a significant change in expression and presumably do not represent false positives due to very low concentration.

**Identification of target pathways.** We consider the set of miRNAs corresponding to each experimental condition as in Fig. 3) and use the online database Diana-MirPath[21] to identify targeted genes and pathways.

**Real Time PCR.** Total RNA has been isolated by NucleoSpin RNA II Kit (Macherey-Nagel, Germany). 1 $\mu$g is reverse transcribed with random hexamers (High Capacity cDNA Archive Kit; Applied Biosystems) in accordance with the manufacturer's instructions. cDNA(5 ng) is amplified in triplicate in a reaction volume of 15 $\mu$l with the TaqMan Gene Expression Assay (Applied Biosystems) and anABI/Prism 7900 HT thermocycler, using a pre-PCR step of 10 minutes at 95 °C, followed by 40 cycles of 15 seconds at 60 °C. and 60 seconds at 60 °C. For any sample the expression level, normalized to the housekeeping gene encoding GAPDH, is determined with the comparative threshold cycle (Ct) method as previously described[54]. Primers used for Real Time PCR are reported by Spagnuolo[54] and in Table S3.

**Western Blot.** Confluent cells were lysed by boiling in a modified Laemmli sample buffer (2% SDS, 20% glycerol, and 125 mM Tris-HCl, pH6.8). Equal amount of proteins were loaded on gel and separated by SDS-PAGE and transferred to a Protran nitrocellulose membrane (Whatman). After blocking, primary and HRP-linked secondary antibodies, specific bindings were detected by chemiluminescence system (GE Healthcare).

**Population dynamics model for miRNA induced phenotypic switching.** We propose a simple model for CSC population dynamics including CSCs $S$, duplicating cancer cells $C$ and non duplicating (senescent) cells $D$. The kinetics for the model is based on the fact a CSC duplicates symmetrically into two CSC with probability $p_2$ or into two cancer cells with probability $p_0$, or asymmetrically into one CSC and one cancer cell with probability $p_1$, such that $p_0 + p_1 + p_2 = 1$. The kinetics of the model only depends on the combination $\varepsilon = p_2 - p_0$, the average relative increase in the number of CSCs after one duplication[31] Cancer cells can switch to CSCs with a probability $p(\mu)$ depending on the concentration of relevant miRNAs $\mu$, or duplicate and give rise to two non-duplicating cancer cells with probability $1 - p$. For simplicity the rates of duplication and switching are taken to be equal to $R_d$ for all the cells yielding:

$$\frac{dS}{dt} = R_d(\varepsilon S + p(\mu)(C + D)), \quad (1)$$

$$\frac{dC}{dt} = R_d(1 - \varepsilon)S - C(1 + p(\mu)), \quad (2)$$

$$\frac{dD}{dt} = R_d(2C - p(\mu)D). \quad (3)$$

When $p = 0$ the model is equivalent to the CSC model introduced in LaPorta et al.[31] (in the limit $M = 1$), while the case of constant $p$ has been studied in[16]. Here we study the case in which the switching probability is not constant but is controlled by the level of relevant miRNAs, which responds to the depletion of the CSC population, as observed experimentally. We assume that the miRNAs regulating phenotypic switching are produced with a rate that rapidly vanishes when the fraction of CSCs $f_S = S/(S + C + D)$ is sufficiently large, and are cleared at constant rate $\gamma$:

$$\frac{d\mu}{dt} = \beta \exp(-f_s/s_0) - \gamma\mu. \quad (4)$$

The switching probability is activated when the level of miRNA is above a threshold $\mu_0$. To avoid artificial singularities, we model the threshold by a smooth sigmoid function

$$p(\mu) = (1 + \tanh((\mu - \mu_0)/\sigma)). \quad (5)$$

Simulations are performed using a custom made python code. Parameters ($R_d = 0.4$ days$^{-1}$, $\varepsilon = 0.05$, $\mu_0 = 1$., $\sigma = 0.03$, $s_0 = 0.005$, $\beta = 1$ days$^{-1}$, $\gamma = 0.015$ days$^{-1}$) are chosen in a way that yields a dynamics similar to the experiments.

**miRNA-222 silencing protocol.** miRNA-222 is silenced by reverse transfection according to the manufacturer's instructions using Validated AMBION SILENCER Select siRNA 222 (4390818, LifeTechnology). As positive control siRNA targeted GAPDH is also used (4390849, AMBION, Life Technology). Lipofectamine RNAiMAX transfection reagent is used to transfect the cells according to Invitrogen protocol. The cells are collected 72 hrs after transfection for further studies.

**Analysis of miRNA and transcriptome of human breast cancer.** We used the METABRIC dataset of human breast tumor that contains mRNA expression data of 1980 tumors and 1283 miRNA expression[47,55]. We also used the ESC-like signature which identified a set of genes from 13 genes sets associated with human ESC identity[56]. We divided the METABRIC patients into ESC-like and non ESC-like tumors. In particular, as described by Ben Porath et al.[56], tumors with overexpression of the ESC exp1 gene set and underexpression of the PRC2 targets genes set were defined as ES-like tumors (418 tumors), while tumors with underexpression of the ESC exp1 set and overexpression of the PRC2 targets set were defined as non ESC-like tumors (666 tumors). 7122 genes were significantly overexpressed in the METABRIC ESC-like tumors, and 5594 genes were significantly underexpressed (1% FDR; 2-samples t-test). Among the METABRIC tumors which were divided into ESC-like and non ESC-like tumors, miRNA expression data was measured on 255 ESC-like and 457 non ESC-like tumors. 156 miRNAs were significantly overexpressed in the ESC-like tumors, and 163 miRNAs were significantly underexpressed (1% false discovery rate; 2-samples t-test).

**Statistical Analysis.** Statistical significance analysis is performed using the Kolmogorov-Smirnov test.

### Acknowledgements

ALS and SZ are supported by the ERC advanced grant SIZEFFECTS. SZ acknowledges support from the Academy of Finland FiDiPro progam, project 13282993. JSP and SZ are supported by NSF and CNR through Materials World Network: Cooperative Activity in Materials Research between US Investigators and their Counterparts Abroad in Italy (NSF DMR 1312160). CRM acknowledges support from NSF grant IOS 1127017. CAMLP is supported by PRIN MIUR 2010 and thanks the visiting professor program of Aalto University.

### Author Contributions

A.L.S. analyzed miRNA data. E.C. performed flow cytometry experiments. S.C. and A.P. performed STR experiments. C.G. performed real time PCR and western blots. N.B.B. analyzed breast cancer data. C.R.M. and J.P.S. assisted in statistical analysis, model design, and writing of the paper. S.Z. designed and simulated the model, analyzed FACS data and assisted in writing of the paper. C.A.M.L.P. designed the project, performed and supervised experiments, and wrote the paper.

### Additional Information

**Supplementary information** accompanies this paper at http://www.nature.com/srep

**Competing financial interests:** The authors declare no competing financial interests.

**How to cite this article**: Sellerio, A. L. *et al.* Overshoot during phenotypic switching of cancer cell populations. *Sci. Rep.* **5,** 15464; doi: 10.1038/srep15464 (2015).